\documentstyle[psfig]{mn}				

\newif\ifAMStwofonts

\def\simlt{\lower.5ex\hbox{$\; \buildrel < \over \sim \;$}}
\def\simgt{\lower.5ex\hbox{$\; \buildrel > \over \sim \;$}}

\ifoldfss
  \ifCUPmtlplainloaded \else
    \NewTextAlphabet{textbfit} {cmbxti10} {}
    \NewTextAlphabet{textbfss} {cmssbx10} {}
    \NewMathAlphabet{mathbfit} {cmbxti10} {} 
    \NewMathAlphabet{mathbfss} {cmssbx10} {} 
  \fi
  \ifAMStwofonts
    \ifCUPmtlplainloaded \else
      \NewSymbolFont{upmath} {eurm10}
      \NewSymbolFont{AMSa} {msam10}
      \NewMathSymbol{\upi}     {0}{upmath}{19}
      \NewMathSymbol{\umu}     {0}{upmath}{16}
      \NewMathSymbol{\upartial}{0}{upmath}{40}
      \NewMathSymbol{\leqslant}{3}{AMSa}{36}
      \NewMathSymbol{\geqslant}{3}{AMSa}{3E}

      \let\leq=\leqslant 
      \let\geq=\geqslant 
    \fi
  \fi
\fi 

\ifnfssone
  \newmathalphabet{\mathit}
  \addtoversion{normal}{\mathit}{cmr}{m}{it}
  \addtoversion{bold}{\mathit}{cmr}{bx}{it}
  \newmathalphabet{\mathbfit} 
  \addtoversion{normal}{\mathbfit}{cmr}{bx}{it}
  \addtoversion{bold}{\mathbfit}{cmr}{bx}{it}
  \newmathalphabet{\mathbfss} 
  \addtoversion{normal}{\mathbfss}{cmss}{bx}{n}
  \addtoversion{bold}{\mathbfss}{cmss}{bx}{n}
  \ifAMStwofonts
    \ifCUPmtlplainloaded \else
      \UseAMStwoboldmath
      \makeatletter
      \new@mathgroup\upmath@group
      \define@mathgroup\mv@normal\upmath@group{eur}{m}{n}
      \define@mathgroup\mv@bold\upmath@group{eur}{b}{n}
      \edef\UPM{\hexnumber\upmath@group}
      \new@mathgroup\amsa@group
      \define@mathgroup\mv@normal\amsa@group{msa}{m}{n}
      \define@mathgroup\mv@bold\amsa@group{msa}{m}{n}
      \edef\AMSa{\hexnumber\amsa@group}
      \makeatother
      \mathchardef\upi="0\UPM19
      \mathchardef\umu="0\UPM16
      \mathchardef\upartial="0\UPM40
      \mathchardef\leqslant="3\AMSa36
      \mathchardef\geqslant="3\AMSa3E

      \let\leq=\leqslant 
      \let\geq=\geqslant 
    \fi
  \fi
\fi 

\ifnfsstwo
  \DeclareMathAlphabet{\mathbfit}{OT1}{cmr}{bx}{it}
  \SetMathAlphabet\mathbfit{bold}{OT1}{cmr}{bx}{it}
  \DeclareMathAlphabet{\mathbfss}{OT1}{cmss}{bx}{n}
  \SetMathAlphabet\mathbfss{bold}{OT1}{cmss}{bx}{n}
  \ifAMStwofonts
    \ifCUPmtlplainloaded \else
      \DeclareSymbolFont{UPM}{U}{eur}{m}{n}
      \SetSymbolFont{UPM}{bold}{U}{eur}{b}{n}
      \DeclareSymbolFont{AMSa}{U}{msa}{m}{n}
      \DeclareMathSymbol{\upi}{0}{UPM}{"19}
      \DeclareMathSymbol{\umu}{0}{UPM}{"16}
      \DeclareMathSymbol{\upartial}{0}{UPM}{"40}
      \DeclareMathSymbol{\leqslant}{3}{AMSa}{"36}
      \DeclareMathSymbol{\geqslant}{3}{AMSa}{"3E}

      \let\leq=\leqslant 
      \let\geq=\geqslant 
    \fi
  \fi
\fi 

\ifCUPmtlplainloaded \else
  \ifAMStwofonts \else 
    \def\upi{\pi}
    \def\umu{\mu}
    \def\upartial{\partial}
  \fi
\fi

\pagerange {1--22}

\title [The accretion disc and gas stream of UX~UMa]
	{HST spatially-resolved spectra of the accretion disc and gas stream of
	the nova-like variable UX~Ursae~Majoris
	\thanks{Based on observations with the
	NASA/ESA {\em Hubble Space Telescope}, obtained at the Space Telescope
	Science Institute, which is operated by the Association of Universities
	for Research in Astronomy, Inc., under NASA contract NAS~5-2655}
	}

\author[R. Baptista et~al.]
	{Raymundo Baptista $^{1,2,3}$, Keith Horne $^2$, Richard A. Wade $^4$, Ivan Hubeny $^5$,
	\cr Knox S. Long $^3$ and Ren\'e G. M. Rutten $^6$ \\
	$^1$~Depto. de F\i{i}sica, Universidade Federal de Sta.\,Catarina,
	Campus Trindade, 88040-900 Florian\'opolis/SC, Brazil, \\ Email: bap@fsc.ufsc.br \\
	$^2$~University of St.\,Andrews, School of Physics \& Astronomy, North Haugh,
	St.\,Andrews, Fife KY16 9SS, Scotland, \\ Email: kdh1@st-andrews.ac.uk \\
	$^3$~Space Telescope Science Institute, 3700 San Martin Drive, Baltimore,
	MD 21218, USA, Email: long@stsci.edu \\
	$^4$~The Pennsylvania State University, Dept.\ of Astronomy and Astrophysics,
	525 Davey Laboratory, University Park, PA 16802, USA \\ Email: wade@astro.psu.edu\\
	$^5$~NASA Goddard Space Flight Center, Greenbelt, MD 20771, Email:
	hubeny@stars.gsfc.nasa.gov \\
	$^6$~~Isaac Newton Group, Apartado de correos 321, E-38780 Santa Cruz de La Palma,
	Spain, Email: rgmr@ing.iac.es \\
}

\date{Submitted 1997 December 21; accepted 1998 March 11}
\pubyear{1998}

\begin{document}

\maketitle

\begin{abstract}

Time-resolved eclipse spectroscopy of the nova-like variable UX~UMa
obtained with the HST/FOS on 1994 August and November is analyzed with
eclipse mapping techniques to produce spatially resolved spectra of its
accretion disc and gas stream as a function of distance from disc centre.

The inner accretion disc is characterized by a blue continuum filled
with absorption bands and lines which cross over to emission with
increasing disc radius, similar to that reported by Rutten et~al (1994)
at optical wavelengths.  The comparison of spatially resolved spectra at
different azimuths reveals a significant asymmetry in the disc emission
at UV wavelengths, with the disc side closest to the secondary star
showing pronounced absorption by an `iron curtain' and a Balmer jump in
absorption.  These results suggest the existence of an absorbing ring of
cold gas whose density and/or vertical scale increase with disc radius.
The spectrum of the infalling gas stream is noticeably different from the
disc spectrum at the same radius suggesting that gas overflows through
the impact point at disc rim and continues along the stream trajectory,
producing distinct emission down to $0.1\; R_{L1}$.

The spectrum of the uneclipsed light shows prominent emission lines
of Ly$\alpha$, N\,{\sc V} $\lambda 1241$, Si\,{\sc IV} $\lambda 1400$,
C\,{\sc IV} $\lambda 1550$, He\,{\sc II} $\lambda 1640$, and Mg\,{\sc II}
$\lambda 2800$, and a UV continuum rising towards longer wavelengths. The
Balmer jump appears clearly in emission indicating that the uneclipsed
light has an important contribution from optically thin gas. The lines
and optically thin continuum emission are most probably emitted in a
vertically extended disc chromosphere + wind.

The radial temperature profiles of the continuum maps are well described
by a steady-state disc model in the inner and intermediate disc regions
($R \leq 0.3 R_{L1}$). There is evidence of an increase in the mass
accretion rate from August to November (from \.{M}$= 10^{-8.3\pm 0.1}
\;{\rm to}\; 10^{-8.1\pm 0.1}\; M_{\odot} \; yr^{-1}$), in accordance
with the observed increase in brightness.  Since the UX\,UMa disc seems
to be in a high mass accretion, high-viscosity regime in both epochs,
this result suggests that the mass transfer rate of UX~UMa varies 
substantially ($\simeq 50$ per cent) on time scales of a few months.

It is suggested that the reason for the discrepancies between the
prediction of the standard disc model and observations is not an
inadequate treatment of radiative transfer in the disc atmosphere, but
rather the presence of additional important sources of light in the system
besides the accretion disc (e.g., optically thin continuum emission from
the disc wind and possible absorption by circumstellar cool gas).

\end{abstract}


\section{Introduction}

Accretion discs are an important phenomenon in astrophysics, invoked to
solve a wide range of astrophysical problems ranging from planetary
formation to quasar energetics (Frank, King \& Raine 1992).  Although
considerable effort in both observation and theory has been invested
over the past decade, the structure and underlying physics of accretion
discs remains poorly understood. Major unsolved problems include the
nature of the viscosity mechanism -- responsible for the spiraling
inward of the disc material -- (the angular momentum problem), the fate
of the kinetic energy expended at the inner edge of the accretion disc
(the boundary layer problem), the vertical structure of the disc (the
Balmer decrement problem), and the outflow of matter in connection with
a disc wind (possibly a solution to, or at least an element of, the
boundary layer problem). Progresses in solving these issues has
been hampered because most of the previous observational
constraints provided only the spectrum of the total light from the
disc.  A better understanding of the physics of accretion discs
requires spatially-resolved studies.

Cataclysmic Variables (CVs) are mass-exchanging binary systems
containing a white dwarf and a late-type star (Warner 1995). If the
white dwarf is not strongly magnetized ($B < 10^{6}$~G) an accretion
disc is formed.  Accretion discs in non-magnetic CVs cover a range of
accretion rates and viscosity states. For example, {\em dwarf novae}
undergo large outbursts ($\Delta m = 3-5$~mag, typical duration of 5-10
days) which reflects changes in the structure of the discs -- from a
cool, optically thin, low viscosity state to a hot, optically thick,
high viscosity state -- and which are usually parameterized as a large
change in the mass accretion rate ( \.{M}$= 10^{-11} \; M_\odot \;
yr^{-1} \mapsto 10^{-9} \; M_\odot \; yr^{-1}$.  See, e.g.  Pringle,
Verbunt \& Wade 1986).  On the other hand, {\em nova-like} variables
seem to be permanently in a high viscosity state, presumably as a
result of the fact that the accretion rate is always high.
Because the nature of the other constituents in these systems -- the
white dwarf and the normal star -- are reasonably well understood,
and because orbital variations often provide considerable insight into
the system geometry, non-magnetic CVs are the ideal laboratories for
understanding accretion discs.  Eclipsing systems are particularly useful
since the occultation of the accretion disc by the late-type star provides
information about the disc's spatial structure through the eclipse shape.

The eclipse mapping method (Horne 1985, 1993; Rutten, van Paradijs \&
Tinbergen 1992; Baptista \& Steiner 1993) assembles the information
contained in the eclipse shape into a map of the disc surface
brightness distribution.  When applied to time-resolved spectroscopy
through eclipses this technique delivers the spectrum of the disc at
any position on its surface.  Information on the radial dependence of
the temperature and vertical temperature gradients (for optically thick
regions), or temperature, surface density and optical depth (where the
disc is optically thin) can be obtained by comparing such spectra with
the predictions of models of the vertical disc structure.  The spatial
structure of the emission-line regions over the disc can be similarly
mapped from data of high spectral resolution.
Furthermore, by studying the time-variations in the structure of accretion
discs of dwarf novae undergoing outbursts it may be possible to uncover
the nature of the (so far unknown) viscosity mechanism which drives
accretion discs.

UX~UMa is a well known, bright ($V \simeq 12.5$) eclipsing nova-like
variable with an orbital period of 4.72~hr. Eclipse mapping in broad-bands
(Horne 1983; Rutten et~al. 1992) shows that its accretion disc is
optically thick and is close to a steady state at a mass accretion rate
of $\simeq 10^{-8} \; M_\odot \; yr^{-1}$.  The broad-band mapping was
extended to spectrally-resolved mapping in the optical range by Rutten et
al. (1993, 1994).  Their results show that the continuum becomes fainter
and redder with disc radius -- reflecting a radial temperature gradient
-- and reveal that the Balmer lines are seen in absorption in the inner
disc but in emission in the outer disc.  Baptista et al. (1995) performed
a similar study using HST data in narrow spectral windows about
the C\,{\sc IV} 1550 and He\,{\sc II} 1640 line regions. This study showed
that the UV continuum reasonably follows the $T \propto R^{-3/4}$ law for
steady mass accretion, confirming the results from the optical analysis.
The C\,{\sc IV} and He\,{\sc II} line profiles are dominated by emission
from the disc wind. Spatially-resolved spectra reveal that these lines
appear as narrow absorption features at disc centre and change with
increasing radius to broad emission in the outer disc regions besides
showing large uneclipsed components.  This behaviour is similar to that
found for the Balmer lines and suggests that these optical lines
may also have a wind component.

In this paper, we report on the ultraviolet (UV) and optical mapping of
the accretion disc and gas stream of UX~UMa, based on observations
made with the {\it Faint Object Spectrograph} (FOS) on the {\it Hubble
Space Telescope} (HST). The reader is referred to Knigge et~al. (1998a)
for an initial description of these observations, with emphasis on
the spectral properties of the integrated spectra of the accretion
disc, the bright spot and the uneclipsed light.  Sect.\,\ref{observa}
describes the data and its reduction. The extraction of narrow-band
light curves and their analysis with eclipse mapping techniques are
described in Sect.\,\ref{analise}. Sect.\,\ref{results} presents and
discusses spatially resolved spectra of the accretion disc and the gas
stream region as a function of distance from disc centre, the spectrum
of the uneclipsed light, and the radial temperature distribution in the
ultraviolet. The possible influence of the assumed eclipse geometry on the
results is addressed in Sect.\,\ref{geo}.  Sect.\,\ref{discuss} discusses
the implications of the results in the context of disc atmosphere models.
The results are summarized in Sect.\,\ref{conclusao}.

\section{Observations} \label{observa}

Time-resolved low-resolution spectroscopy centred on eclipses of
UX\,UMa was obtained with HST/FOS in the `rapid readout' mode on 1994
August (G160L, $1100-2500$~\AA, spectral resolution of $\Delta\lambda =
3.5$~\AA\ pixel$^{-1}$) and 1994 November (PRISM, $1600-8500$ \AA, spectral
resolution varying from 1~\AA\ pixel$^{-1}$ near the short wavelength end
to more than 200~\AA\ pixel$^{-1}$ at the long wavelength end), at a time
resolution of 5.3\,s, in a total of 4 data sets and 3122 spectra. The
nominal sampling of the FOS spectra was increased by a factor 2 by
magnetically deflecting the light beam along the dispersion direction
by 1/2 of the diode width in order to make the array of diodes sample
a different set of wavelengths, and by repeating the process until the
spectra has shifted by an integer number of diodes (a procedure called
`sub-stepping').  The total exposure per pixel in each spectrum
is 2.59\,s.  The G160L observations were taken with the $0.86\arcsec$
square aperture to reduce the contribution of the geocoronal Ly$\alpha$
emission, while the PRISM observations were obtained with a larger
$3.66\arcsec \times 3.71\arcsec$ square aperture. The observations are
summarized in Table\,1.
%
%
\begin{table*}
\centering
\begin{minipage}{140mm}
\caption{Journal of the observations.}
\begin{tabular}{@{}clccccclc@{}}
\hline
Run & \multicolumn{1}{c}{Date} & \multicolumn{2}{c}{UT} & Phase & Eclipse &
No. of & Instrument & Scaling \\ [-0.3ex]
    &      &    Start  &     End    & range &  cycle  & spectra && factor \\
\hline
1 & 1994 Aug 3  & 02:52:33 & 03:54:32 & $-0.090,+0.130$ & 28793 & 691 &
FOS/G160L & 1.00 \\
2 & 1994 Aug 3  & 07:43:06 & 08:45:05 & $-0.063,+0.156$ & 28794 & 691 &
FOS/G160L & 1.05 \\
3 & 1994 Nov 24 & 04:33:46 & 05:51:49 & $-0.156,+0.120$ & 29368 & 870 &
FOS/PRISM & 1.00 \\ 
4 & 1994 Nov 24 & 18:55:04 & 20:13:07 & $-0.115,+0.161$ & 29371 & 870 &
FOS/PRISM & 0.96 \\
\hline
\end{tabular}
\end{minipage}
\end{table*}

The observations were reduced using procedures similar to the standard
STSDAS pipeline, and included flat-field and geomagnetically induced
motion (`GIMP') corrections, background and scattered light
subtraction, wavelength and absolute flux calibrations.
Further details of the reduction procedures are given
in Knigge et~al. (1998a).

The top panel in Fig.\,\ref{ffig1} shows light curves of the four HST runs
in the overlapping wavelength region (1700-2300~\AA) while the lower panel
shows the corresponding light curves for runs 3 and 4 in a narrow band in
the optical.
%
%
UX\,UMa was brighter on November than on August, by up to 70 per cent
prior to eclipse. The November UV light curves show a pronounced
decrease in flux level from before to after eclipse -- interpreted by
Knigge et~al.  (1998a) as due to anisotropic emission from the bright
spot -- which is not conspicuous in the optical range.  The UV
continuum light curves show significant flickering activity outside of
eclipse, relatively much stronger than observed in the optical range.
The UV eclipses are deep and asymmetric, showing an egress shoulder,
however no clear evidence of a compact bright spot is seen.
Quasi-periodic oscillations of period 25-30\,s are easily seen in the
November light curves. The analysis of these oscillations is the
subject of a separate paper (Knigge et~al. 1998b).

\begin{figure*}
 \centerline{\psfig{figure=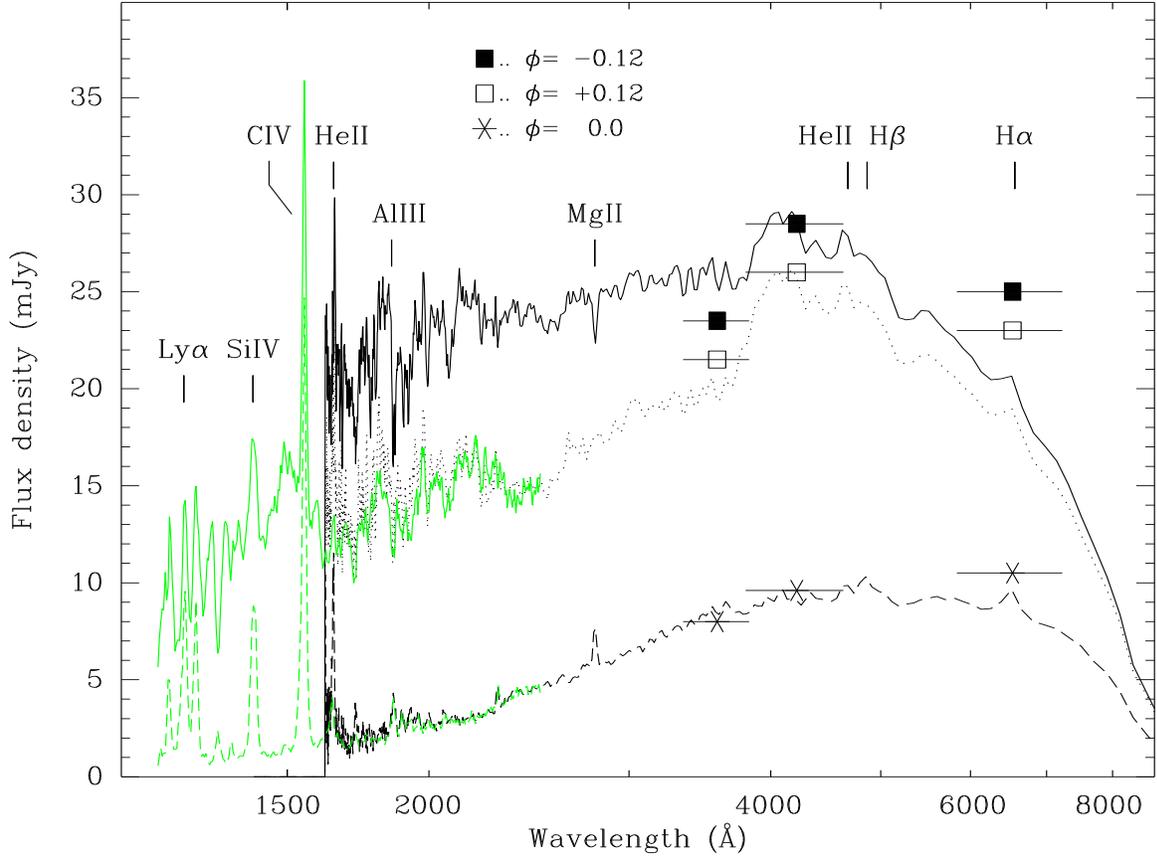,angle=-90,width=19cm,rheight=14cm}}
	\caption{ Average spectra for the Aug 94 (light gray) and Nov 94 (black)
	data. Average spectra prior to eclipse (phase range $\phi= -0.09$ to
	$-0.06$ for August and $-0.164$ to $-0.053$ for November) are shown as
	solid lines whereas average mid-eclipse spectra
	(from $\phi= -0.01$ to $+0.01$) are shown as dashed lines. For the Nov 94
	data an average spectrum after eclipse (from $\phi= +0.083$ to $+0.11$)
	is also shown as a dotted line.
	Average UBR photometry (from Horne 1983) prior to eclipse ($\phi=-0.12$,
	filled squares), after eclipse ($\phi=+0.12$, open squares) and at
	mid-eclipse phases ($\phi=0.0$, asterisks) are shown for comparison.
	Horizontal bars indicate the FWHM of these passbands. }
	\label{ffig2}
\end{figure*}

Figure\,\ref{ffig2} displays average spectra for the August (light gray)
and November (black) data at selected phases. Average spectra prior to
eclipse are shown as solid lines whereas average mid-eclipse spectra
are shown as dashed lines. For the November data, an average spectra after
eclipse is also shown as a dotted line.
%
%

The sharp decrease in flux at the red end of the November spectra is
a distortion caused by an unsolved calibration problem affecting the
PRISM data, most probably due to the combination, at these wavelengths,
of a large and highly non linear dispersion with a sharp decrease in
diode sensitivity.  UBR photometry prior to eclipse, after eclipse and
at mid-eclipse phases derived from the light curves of Horne (1983)
are plotted in Fig.\,\ref{ffig2} for comparison.  The fluxes of the
PRISM spectra are consistent with the photometric measurements at U and
B and are underestimated by $\simeq 20$ per cent at R, the distortion
increasing for longer wavelengths.

The comparison of the spectra prior to eclipse of the August and the
November data underscores the observed differences in the brightness of
UX~UMa at these epochs.  The November spectrum shows a substantially
higher continuum level at the overlapping wavelength region, and the
absorption features are more pronounced than in the August data.  The
He\,{\sc II} $\lambda 1640$ line -- hardly seen in the August spectrum --
appears strongly in emission.  Although the spectrum prior to eclipse
of August is a reasonable match to the spectrum after eclipse of November,
the He\,{\sc II} line is considerably stronger in the later.

The UV out-of-eclipse spectrum shows prominent emission lines (C\,III
$\lambda 1176$, Ly$\alpha$, N\,{\sc V} $\lambda 1240$, Si\,{\sc IV}
$\lambda 1400$, C\,{\sc IV} $\lambda 1550$) as well as many absorption
features and possibly broad absorption bands, particularly near 1900~\AA
~and 2400~\AA.  At mid-eclipse the continuum flux is reduced by a factor
$\simgt 3$, while the emission lines are much less deeply eclipsed and
some of the lines that are seen in absorption in the out-of-eclipse
spectra appear in emission (e.g., He\,{\sc II} $\lambda 1640$).

The reader is referred to Knigge et~al (1998a) for a detailed
analysis of the integrated spectra of UX~UMa.

\section{Data analysis} \label{analise}

\subsection{Light curve construction}

The calibrated spectra were divided into 59 (G160L) and 127 (PRISM)
passbands of 15-30 \AA\ wide in the ultraviolet continuum and $\sim 2300$
km s$^{-1}$ for the stronger UV emission lines.  In the optical ($\lambda
\geq 3600$~\AA) each passband corresponds to a single pixel and therefore
its width increases with wavelength from about 50 \AA\ at 3600 \AA\ to 230
\AA\ at the longer wavelength end.  For each passband a light curve was
constructed by computing the average flux on the corresponding wavelength
range and phase folding the resulting data according to the ephemeris
of Baptista et~al. (1995). A phase correction of $-0.002$ cycles was
further applied to the data to make the centre of the white dwarf eclipse
coincident with phase zero.  For those passbands including emission lines
the light curves comprise the total flux at the corresponding bin with
no subtraction of a possible continuum contribution.

\begin{figure*}
 \centerline{\psfig{figure=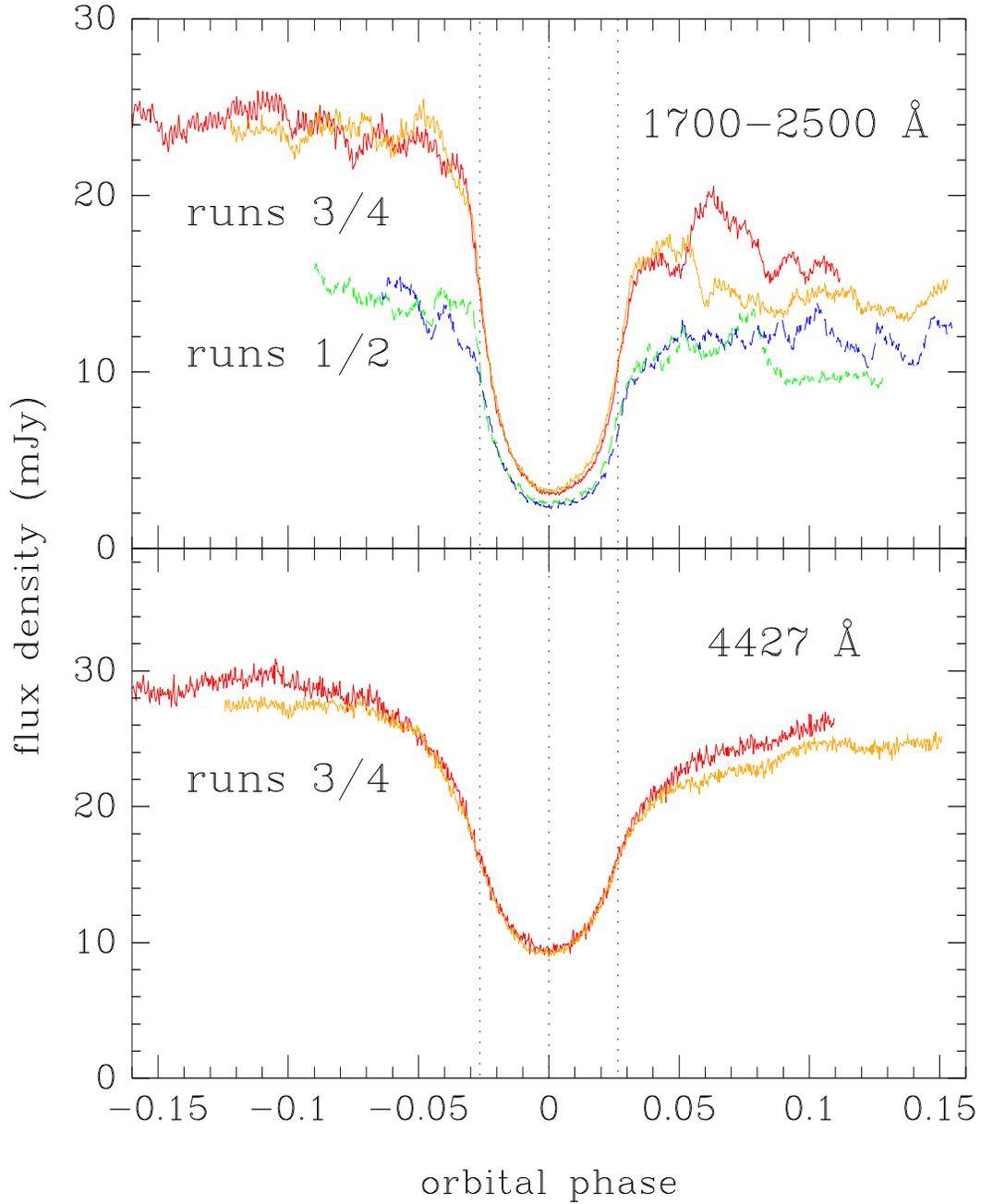,width=17cm,rheight=21cm}}
	\caption{ Top: Light curves of the four HST runs at the same wavelength
	range. Bottom: Light curves for runs 3 and 4 at $\lambda 4427$.
	Vertical dotted lines mark ingress/egress phases of the white dwarf
	and mid-eclipse. }
	\label{ffig1}
\end{figure*}

Small differences in brightness between the runs of each epoch (at the
level of $\leq 5$ per cent) were removed by scaling the light curves to a
common absolute flux level outside eclipse with a wavelength independent
factor, listed in Table\,1.  The light curves in Fig.\,\ref{ffig1}
were scaled by these factors before plotting.

Average light curves were constructed for each passband by combining
the individual light curves, dividing the data into phase bins of
0.002~cycles and computing the median for each bin. The median of the
absolute deviations with respect to the median was taken as the
corresponding uncertainty.
Run 4 shows a pronounced flare centred on phase $\simeq +0.07$ cycles at
UV wavelengths (Fig.\,\ref{ffig1}). The uncertainties around this phase
were artificially increased to reduce the relative
influence of the flare on the shape of the average light curves.

Out-of-eclipse brightness changes are not accounted for by the eclipse
mapping method, which assumes that all variations in the eclipse light
curve are due to the changing occultation of the emitting region by the
secondary star.  Orbital variations in the average profiles were
therefore removed by fitting a spline function to the
phases outside eclipse, dividing the light curve by the fitted
spline, and scaling the result to the spline function value at phase
zero. This procedure removes orbital variations outside eclipse with
only minor effects on the eclipse shape itself.

\subsection{Eclipse mapping} \label{mem}

The eclipse mapping method was used to solve for a map of the disc
brightness distribution and for the flux of an additional uneclipsed
component in each passband.  For the details of the method the reader
is referred to Horne (1985, 1993), Baptista \& Steiner (1993),
Rutten et al. (1994) and Baptista, Steiner and Horne (1996).

For our eclipse map, we adopted a $41 \times 41$ pixel grid centred on
the primary star with side $2 \:R_{L1}$, where $R_{L1}$ is the
distance from the disc centre to the inner Lagrangian point. This
choice provides maps with a nominal spatial resolution of $0.049 \:
R_{L1}$.  The eclipse geometry is specified by the mass ratio $q$
and the inclination $i$. We adopted the parameters of Baptista et~al.
(1995), $i= 71\degr$ and $q=1.0$.
The specific intensities in the eclipse map were computed assuming
R$_{L1}= 0.7 \; R_\odot$ and a distance of 345 pc (Baptista et~al. 1995).

\begin{figure*}
 \centerline{\psfig{figure=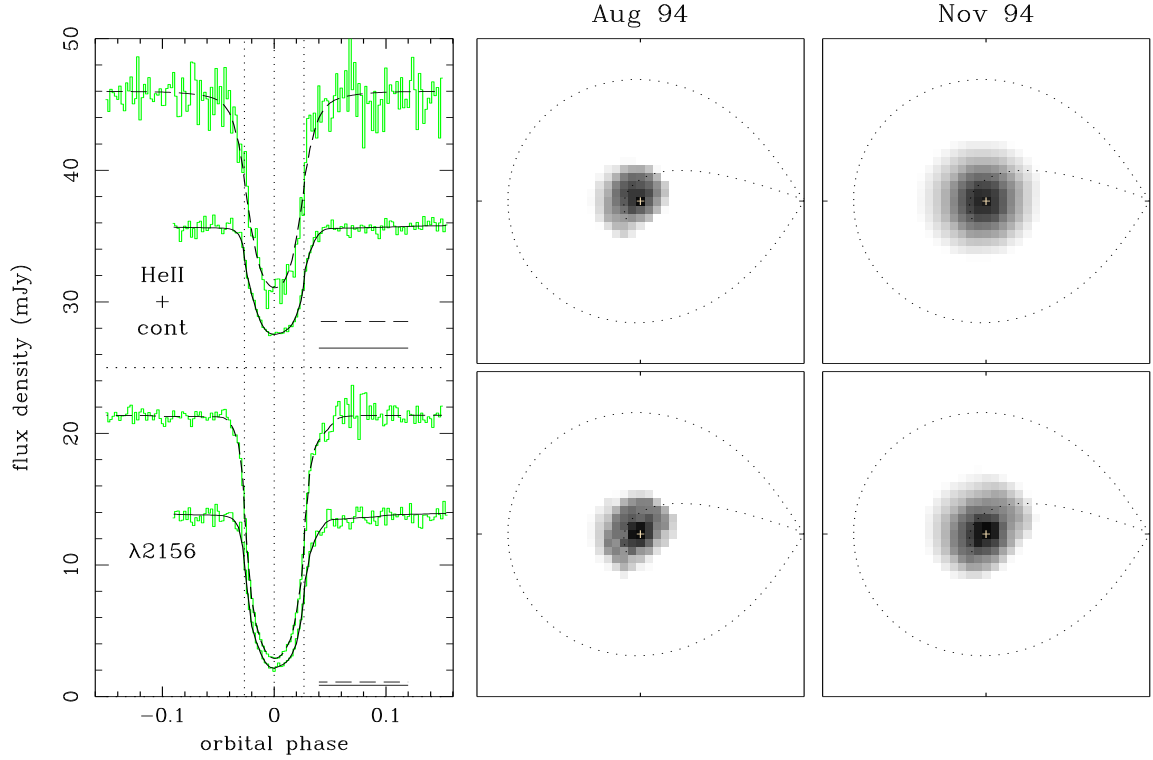,angle=-90,width=18cm,rheight=14cm}}
	\caption{ Changes in the UX~UMa disc structure from Aug 94 to Nov 94.
	Left: eclipse light curves (light gray) and fitted models [solid (Aug)
	and dashed (Nov) lines] for selected passbands. The He\,{\sc II} passband
	includes both line emission and underlying continuum. A horizontal
	dotted line show the zero level for the upper curves. Horizontal solid
	and dashed lines indicate the corresponding uneclipsed components in
	each case. Right: eclipse maps on a logarithmic grayscale.
	White corresponds to $\log I_\nu= -6.35$, and black to
	$\log I_\nu= -2.1$. The notation is the same as in Fig.\,3. }
	\label{ffig4}
\end{figure*}

Average light curves, fitted models, and eclipse maps at selected passbands
are show in Figs.\,\ref{ffig3} and \ref{ffig4}. These will be discussed in
detail in section\,\ref{results}.
%
%

%

\section{Results} \label{results}

\subsection{Disk structure} \label{maps}

In this section we compare eclipse maps at selected passbands in order
to study the structure of the accretion disc at different wavelengths as
well as to assess changes in disc structure related to the significant
increase in brightness level from the August to November observations.

\begin{figure*}
 \centerline{\psfig{figure=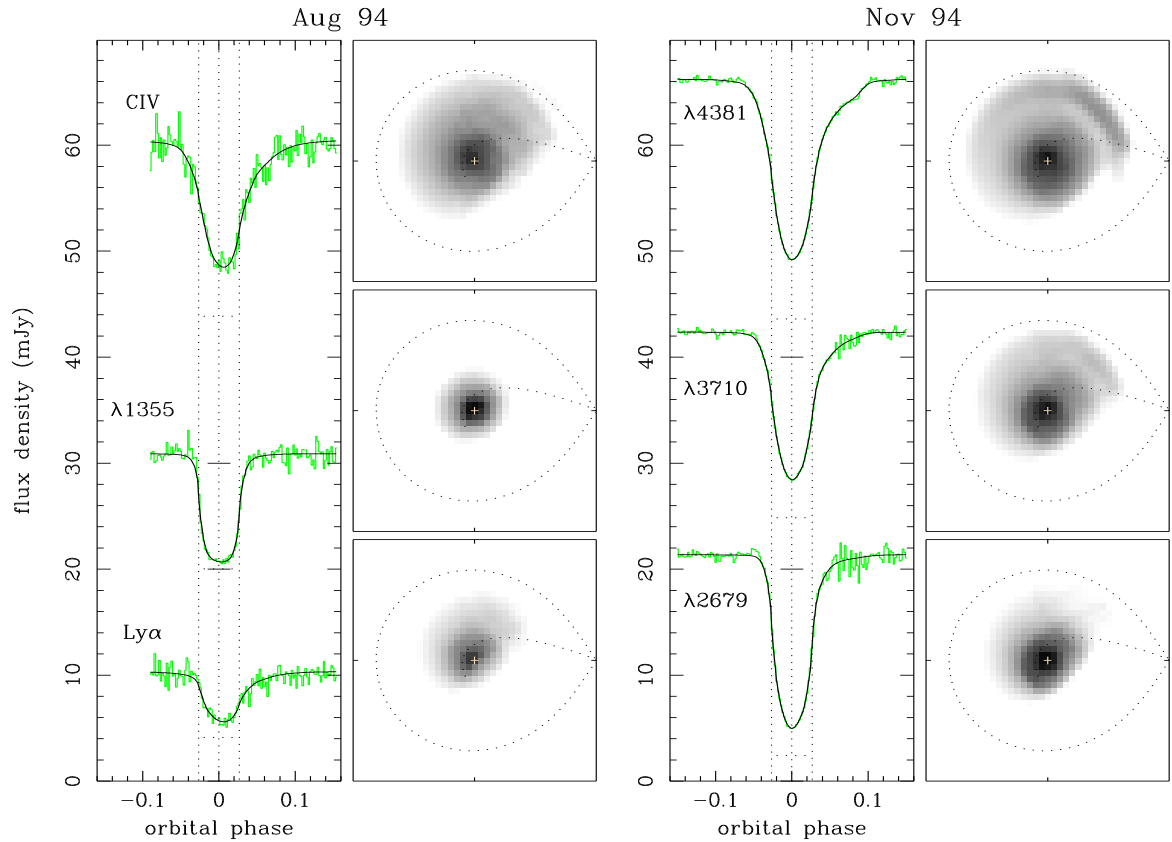,angle=-90,width=18cm,rheight=14cm}}
	\caption{ Light curves and eclipse maps at selected passbands for the
	Aug 94 (left panel) and Nov 94 (right panel) data. Wavelength increases
	upwards. The data light curves are shown as gray histograms and the fitted
	models appear as solid black lines. Solid horizontal bars at mid-eclipse
	show the zero level for the upper curves while dotted horizontal bars
	indicate the uneclipsed component in each case. The C\,{\sc IV} and
	Ly$\alpha$ passbands contain both line emission and underlying continuum.
	The corresponding eclipse maps are shown to the right, in the same
	logarithmic grayscale. Dark regions are brighter; white corresponds
	to $\log I_\nu= -6.7$, and black to $\log I_\nu= -2.3$. Dotted
	curves show the projection of the primary Roche lobe onto the orbital
	plane and the theoretical gas stream trajectory; the secondary star is
	to the right of each panel and the stars rotate counter-clockwise. }
	\label{ffig3}
\end{figure*}

Figure\,\ref{ffig3} shows light curves and eclipse maps at selected
passbands for the August and November data. The panel on the left
compares the disc structure in Ly$\alpha$ and the line centre of
C\,{\sc IV} $\lambda 1550$ with that of the adjacent continuum region
at $\lambda 1355$.  We remark that the Ly$\alpha$ and the C\,{\sc IV}
light curves include the total flux at the corresponding wavelength range,
i.e., we have not subtracted an interpolated continuum.
The differences are obvious. The eclipse in the continuum light curve is
deep and symmetric about mid-eclipse, producing a symmetric brightness
distribution sharply concentrated towards disc centre with no sign of a
bright spot or enhancements in brightness along the gas stream trajectory.
The accretion disc is small and fills only $\simeq 30$ per cent of the
primary lobe.  The light curves of the emission lines show a shallow
eclipse with a clear asymmetry in their egress side. This shape maps into
a brightness distribution which is considerably flatter and more extended
than that of the continuum and which displays a pronounced asymmetry
in the trailing lune of the disc.  These results reveal that the bright
spot and gas stream region contributes a non-negligible fraction of the
flux of these emission lines, in accordance with a suggestion made by
Baptista et~al. (1995).

The right panel of Fig.\,\ref{ffig3} compares light curves and eclipse
maps of the November data at three continuum passbands.  The light
curves show asymmetric egress shoulders which are more pronounced and
end at later phases for longer wavelengths.  The corresponding eclipse
maps show asymmetric structures in the trailing lune of the disc whose
light centre is located at larger disc radii for longer wavelengths.
This suggests that enhanced emission is produced along the gas stream
trajectory {\em downstream} the bright spot position at disc edge, and
that the light centre of this enhanced emission moves inward in radius
for shorter wavelengths. We will further explore this possibility in
section\,\ref{stream}.

Figure\,\ref{ffig4} compares light curves and eclipse maps of the
August and November data at the same two passbands, and is illustrative
of the changes in disc structure caused by the increase in brightness
level at the later epoch. As above, the He\,{\sc II} light curve
includes the total flux at the corresponding wavelength range, with no
subtraction of an interpolated continuum contribution.  The August maps
of both the He\,{\sc II} and the continuum at $\lambda 2156$ show a
small disc with a weak asymmetry in its trailing lune at a position
consistent with the location of the bright spot at the intersection of
the gas stream trajectory with the disc edge.  These maps are also
asymmetric in the sense that the emission from the disc side
farthest away from the inner Lagrangian point (hereafter referred as
`the back side' of the disc) is more pronounced than that of the
side closest to the L1 point (hereafter referred as `the front
side' of the disc).  The similarities of the brightness distribution at
both passbands suggest that the emission at He\,{\sc II} is dominated
by contribution from the underlying continuum and, therefore, that
little or no net He\,{\sc II} line emission exists at this epoch.
In November, the brightness distribution of the map at $\lambda 2156$
has increased in radius and the bright spot is clearly seen at a larger
radial distance than in August. The brightness distribution at He\,{\sc II}
has evolved toward a symmetric and extended distribution significantly
larger in radius than observed in August, with no evidence of a bright
spot. The November He\,{\sc II} distribution is probably dominated by light
from the disc chromosphere + wind, whose emission seems very sensitive
to the brightness level of the system, i.e., to the mass accretion rate
(see section\,\ref{trad}). This supports a previous suggestion made
in this sense by Baptista et~al. (1995).

\subsection{Spatially-resolved disc spectra} \label{back}

Each of the eclipse maps yields spatially-resolved information about
the emitting region on a specific wavelength range.  By combining all
narrow-band eclipse maps we are able to isolate the spectrum of the
eclipsed region at any desired position (e.g., Rutten et al. 1993).

\begin{figure*}
 \centerline{\psfig{figure=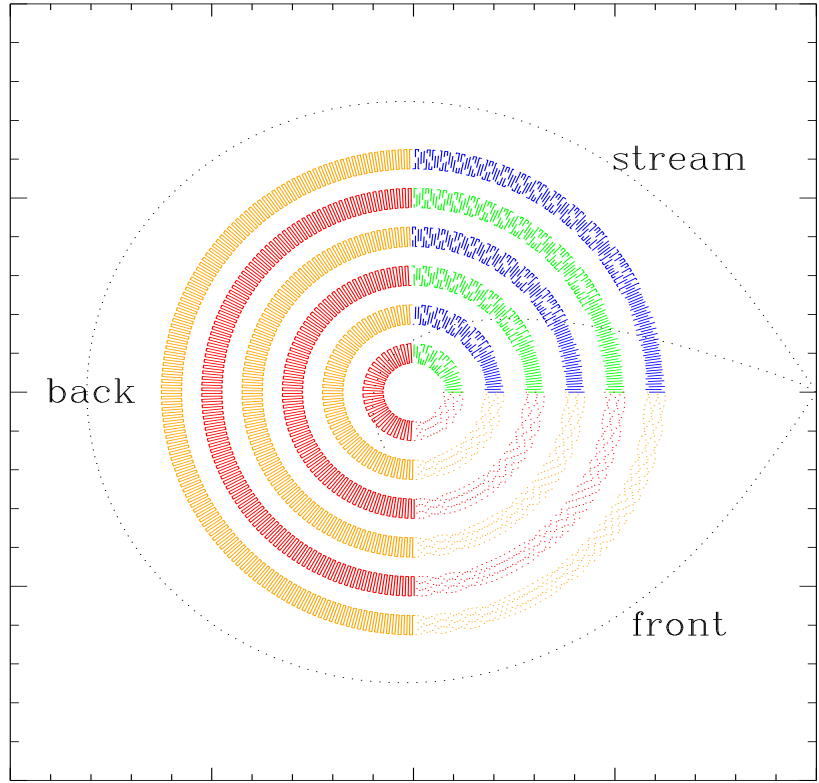,angle=-90,width=18cm,rheight=14cm}}
	\caption{ Annular regions used to extract spatially resolved spectra.
	The disc is divided into three major azimuthal regions (the back side,
	the front side, and the quarter section containing the gas stream
	trajectory), and into a set of 6 concentric annuli with radius increasing
	in steps of $0.1\; R_{L1}$ and of width $0.05\; R_{L1}$ (where $R_{L1}$
	is the distance from disc centre to the inner Lagrangian point). }
	\label{ffig5}
\end{figure*}

Motivated by the observed asymmetries in the disc emission discussed
in Sect.\,\ref{maps} we separated the disc into three major azimuthal
regions:  the back side, the front side, and the region containing the
gas stream trajectory. For each of these regions, we divided the maps
into a set of 6 concentric annuli centred on the white dwarf of width
$0.05\: R_{L1}$ and with radius increasing in steps of $0.1\: R_{L1}$.
The regions used to extract disc spectra are shown in Fig.\,\ref{ffig5}.
%
%
To evaluate the emission from the very centre of the disc, we also
extracted spectra for a circle of width $0.05\: R_{L1}$ centred on
the white dwarf, which includes the central pixel and its four closest
neighbor pixels.

Each spectrum is obtained by averaging the intensity of all pixels
inside the corresponding annulus.  The statistical uncertainties
affecting the average intensities were estimated with a Monte Carlo
procedure (e.g., Rutten et al. 1992; Baptista et~al. 1995).  For a
given narrow-band light curve a set of 10 artificial light curves is
generated, in which the data points are independently and randomly
varied according to a Gaussian distribution with standard deviation
equal to the uncertainty at that point.  These light curves are fitted
with the eclipse mapping algorithm to produce a set of randomized
eclipse maps.  Average intensity values for each of the annular regions
are then computed for all the randomized maps.  The standard deviation
of the average intensities of the randomized maps is taken as the
corresponding uncertainty for each annulus.  The procedure is repeated
for all the narrow-band light curves.

Spatially resolved spectra of the back side region for August and
November are shown in Fig.\,\ref{ffig6}. The lower panel shows the
spectrum of the uneclipsed component. This will be discussed in
detail in Sect.\,\ref{fbg}.
The inner disc in the UV is characterized by a blue continuum filled with
absorption lines and bands which cross over to emission with increasing
disc radius, in accordance with the findings of Rutten et~al. (1994) in
the optical and Baptista et~al. (1995) in the ultraviolet.  The change
in the slope of the continuum with increasing disc radius reflects the
temperature gradient in the accretion disc, with the effective temperature
decreasing outwards.  The comparison of the 1994 Aug and Nov eclipse maps
shows a significant ($\simeq 50$ per cent) increase in brightness at
the later epoch, which is more pronounced at the inner disc region.

At disc centre, the Ly$\alpha$, Si\,{\sc IV} and C\,{\sc IV} lines
show narrow absorption cores superposed on broad absorption wings.
The He\,{\sc II} line -- which is barely seen in the August spectra --
becomes conspicuous in the November spectra with a behaviour similar to
the other strong UV lines.  The Balmer jump appears weakly in absorption
in the inner and intermediate disc and is seen strongly in absorption
at large disc radii. Together with the deep absorption lines seen at the
inner disc, this suggest that the accretion disc of UX~UMa is everywhere
optically thick.
%
%

\subsection{Radial temperature profiles and changes in mass accretion rate}
\label{trad}

The simplest way of testing theoretical disc models is to convert the
intensities in the eclipse maps to blackbody brightness temperatures,
which can then be compared to the radial run of the effective temperature
predicted by steady state, optically thick disc models.  However, one
should bear in mind that the brightness temperature is not a well defined
physical quantity to be compared to the predictions of the standard
disc theory, since the latter formulates the radial profile of the {\em
effective}\ temperature, which measures the total energy dissipated
in the disc per unit surface area.  A relation between the effective
temperature and a monochromatic brightness temperature is non-trivial,
and can only be properly obtained by constructing self-consistent models
of the vertical structure of the disc. Therefore, our analysis here
is meant as preliminary, and should be complemented by detailed disc
spectrum modeling in a future paper.

\begin{figure*}
 \centerline{\psfig{figure=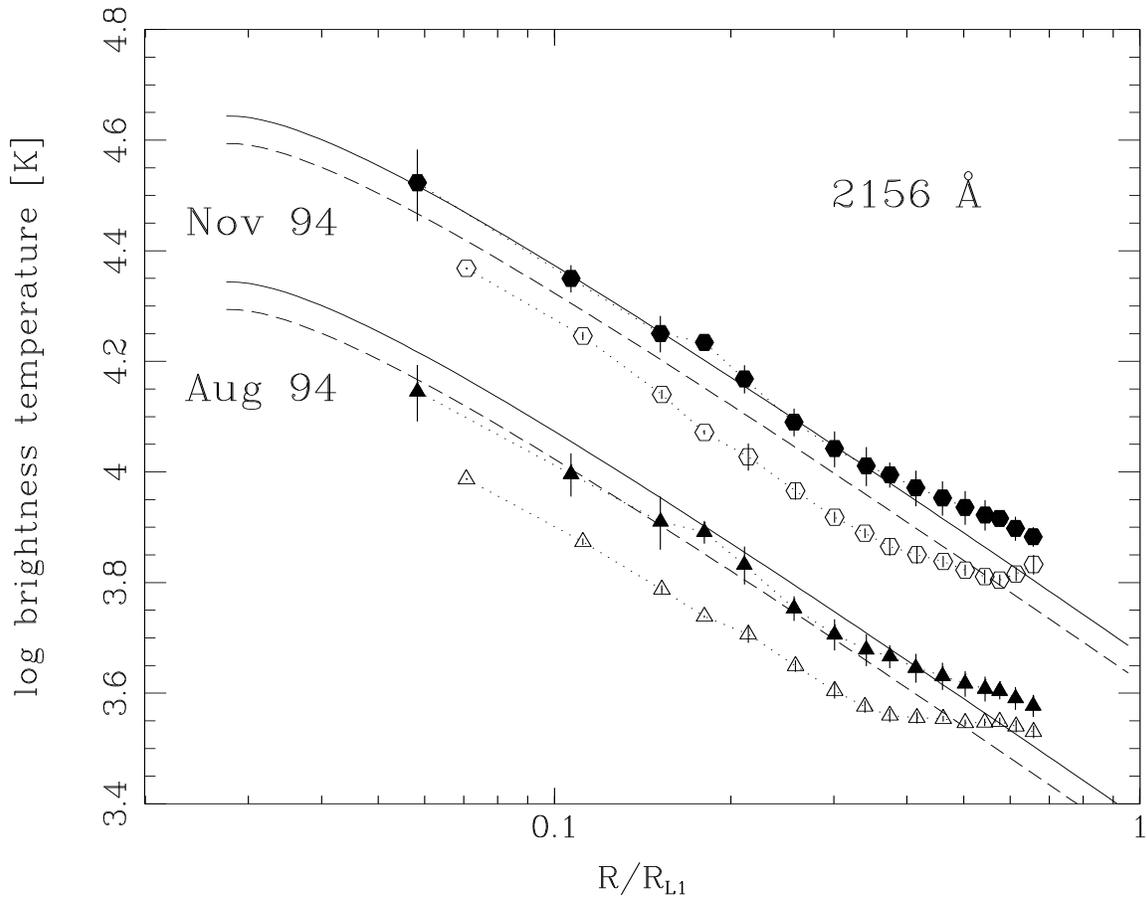,angle=-90,width=19cm,rheight=15cm}}
	\caption{ Radial temperature profiles of the UX~UMa accretion disc at
	$\lambda 2156$.  Steady-state disc models for mass accretion rates of
	$10^{-8.1}$ (solid) and $10^{-8.3}$ (dashed) $M_\odot \; yr^{-1}$
	are plotted for comparison. These models assume $M_1= 0.47 \; M_\odot$
	and $R_1= 0.014 R_\odot$ (Baptista et~al. 1995). The lowest curves are in
	the true temperature scale. The other diagrams were vertically displaced
	by 0.3 dex. Abscissas are in units of the distance from the disc centre
	to the inner Lagrangian point~($R_{L1}$). Filled symbols correspond to
	the profile for the back of the disc, and open symbols correspond to the
	front of the disc. }
	\label{ffig7}
\end{figure*}

Radial brightness temperature profiles for August and November are
plotted in a logarithmic scale in Fig.\,\ref{ffig7}, for a passband
centred at $\lambda 2156$ \AA.  Each temperature shown is the blackbody
brightness temperature that reproduces the observed surface brightness
at the corresponding pixel assuming a distance of 345~pc to UX~UMa
(Baptista et~al. 1995).  Steady-state disc models for mass accretion
rates of $10^{-8.1}$ and $10^{-8.3}\; M_\odot \; yr^{-1}$ are plotted
for comparison, respectively, as solid and dashed lines. These models
assume M$_1= 0.47 \; M_\odot$ and $R_1= 0.014 \; R_\odot$ (Baptista
et~al. 1995). The distribution and comparison curves for November were
vertically displaced by 0.3 dex for a better visualization. Filled
symbols depict the distribution of the back side of the disc, while
open symbols corresponds to the distribution of the front side of the
disc. The temperatures estimated from the front side are systematically
lower than those of the back side, reflecting the fact that the back
side has systematically higher fluxes than the front side.  This may be
due to absorption by cool gas at the disc rim or above the disc plane,
the effect affecting preferably the emission from the front side of
the disc. We will return to this point in Sect.\,\ref{front}. For the
purpose of this section, we will only compare the radial temperature
distributions of the back side of the disc.

Brightness temperatures range from $\sim 8000$\,K in the outer disc
($R= 0.6 \; R_{L1}$) to $\simeq 28000$\,K near the white dwarf at disc
centre ($R= 0.06 \; R_{L1}$) in the August map, and from $\sim 8000$\,K
to $\simeq 33000$\,K in the November map, in reasonable accordance with
the previous results of Rutten et~al. (1994) and Baptista et~al. (1995).
The radial temperature profiles of the continuum maps are in good
agreement with the $T \propto R^{-3/4}$ law expected for steady mass
accretion in the inner and intermediate disc regions ($R \leq 0.3 \;
R_{L1}$).  Comparison of continuum maps at different passbands show that
in this region the radial temperature profiles are consistent within
the uncertainties and can be well described by the same disc model.
This underscores the conclusion that the inner and intermediate accretion
disc in UX~UMa consists of opaque thermal radiators and, therefore,
that the brightness temperatures of the eclipse maps may be a good
first approximation to the effective temperatures of the disc surface in
this region.  The brightness temperature profiles become flatter in the
outer disc regions ($R> 0.4 \; R_{L1}$) yielding temperatures which are
higher than predicted by the steady state disc model and which increase
for shorter wavelengths
\footnote{Brightness temperatures at the blue and red end of the G160L
and PRISM data for $R= 0.6 \; R_{L1}$ differ, respectively, by 15 and
25 per cent.},
showing that blackbody emission is not a good approximation to
the spectra in the outer disc region.  This effect is possibly connected
with the absorption by cool gas in the outer disc, as will be discussed
in section\,\ref{front}.

Mass accretion rates were estimated by computing the best fit steady
state disc model in the range $0.1 \leq R/R_{L1} \leq 0.3$. The quoted
uncertainty includes the uncertainty in the intensities of the individual
pixels as computed with the Monte Carlo procedure (Sect.\,\ref{back})
and the dispersion of the data points about the best fit model for a set
of continuum maps covering the spectral region for which the G160L and
PRISM data overlaps ($\lambda\lambda 1600-2500$ \AA). There is evidence
of an increase in the mass accretion rate from August to November (from
\.{M}$= 10^{-8.3\pm 0.1} \;{\rm to}\; 10^{-8.1\pm 0.1}\; M_\odot \;
yr^{-1}$, at the 95 per cent confidence level), in accordance with the
observed increase in brightness.  Since the UX\,UMa disc seems to be
in a high mass accretion, high-viscosity regime in both epochs, this
result suggests that the {\em mass transfer rate} of UX~UMa varies by a
substantial amount ($\simeq 50$ per cent) on time scales of a few months.
%

\subsection{Front-back asymmetries in disc emission} \label{front}

\begin{figure*}
 \centerline{\psfig{figure=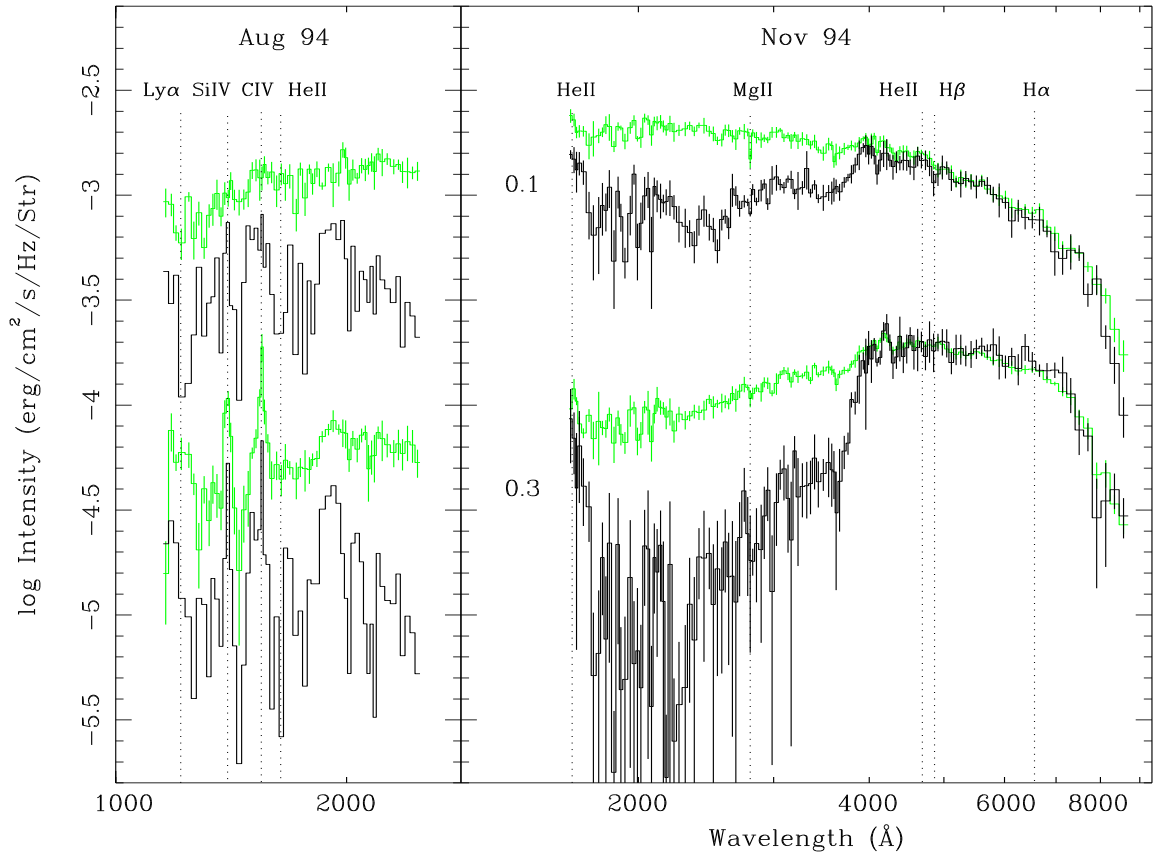,angle=-90,width=17cm,rheight=15cm}}
	\caption{ Comparison of spatially resolved spectra of the back side
	(light gray) and the front side (black) of the disc at two different
	annuli (labels in units of $R_{L1}$). Error bars were computed from
	Monte Carlo simulations with the light curves. To provide a cleaner
	display, error bars for the spectra of the front side in the August
	data were omitted. }
	\label{ffig8}
\end{figure*}

The comparison of the front and back side spectra at the same radius
reveals a significant asymmetry in the disc emission at UV wavelengths,
with the front side of the disc exhibiting a Balmer jump in absorption
and pronounced UV absorption bands -- probably due to a large number
of blended lines of Fe\,II and other species -- present both in the
August and the November observations (Fig.\,\ref{ffig8}). This effect
is reminiscent of that observed previously in OY Car, where the white
dwarf emission seems veiled by an `iron curtain', and was attributed
to absorption by cool circumstellar material, possibly in the outer disc
(Horne et~al. 1994).  We remark that no asymmetry is observed at optical
wavelengths, in accordance with the results of Rutten et~al. (1994).

\begin{figure*}
 \centerline{\psfig{figure=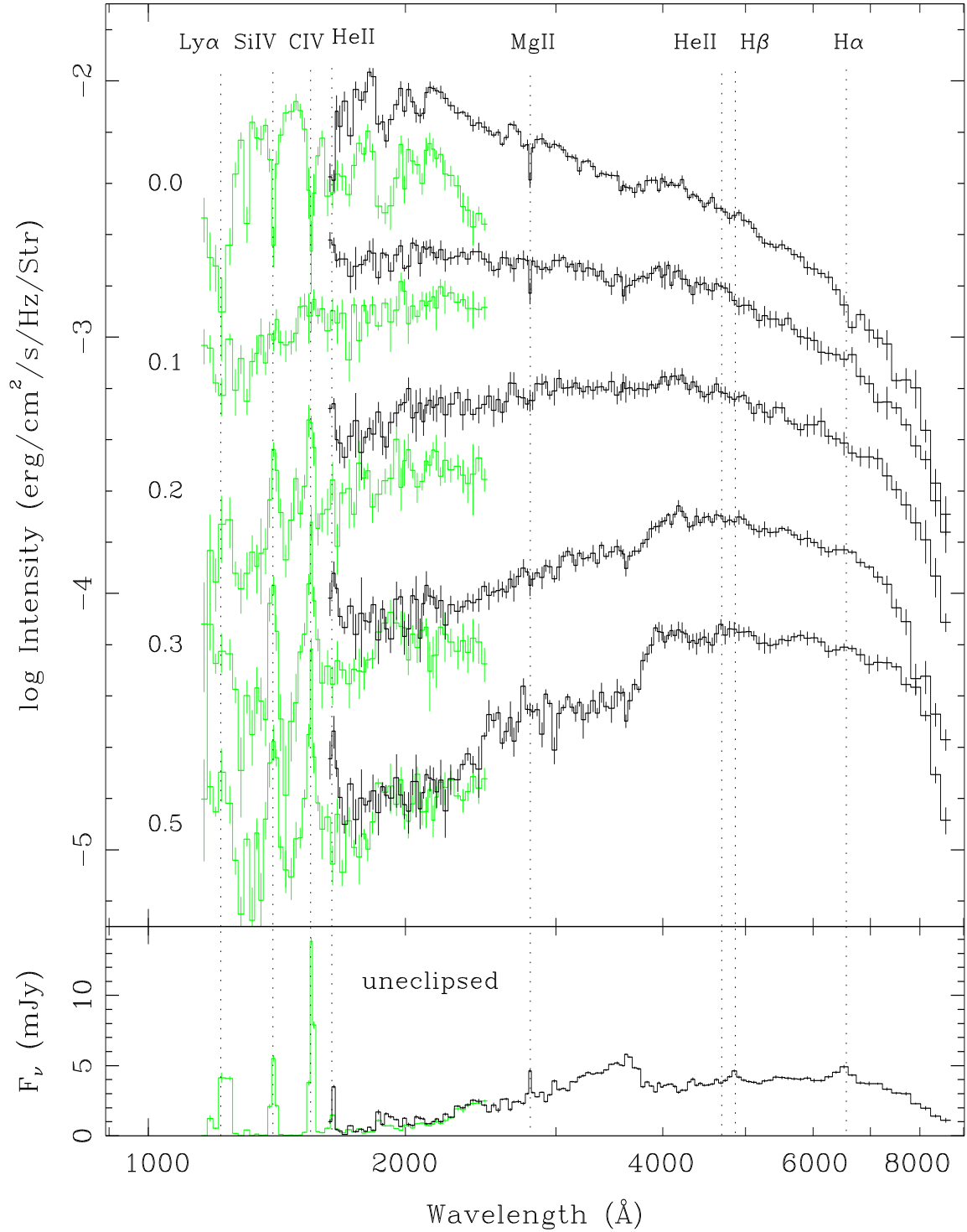,width=17cm,rheight=21cm}}
	\caption{ Spatially resolved spectra of the back region of the UX\,UMa
	accretion disc for the Aug 94 (light) and Nov 94 (dark) data. The
	spectra were computed for a set of concentric annular sections
	(central radius indicated on the left, in units of $R_{L1}$).
	The lower panel shows the spectrum of the uneclipsed light.
	The most prominent line transitions are indicated by vertical dotted
	lines.  Error bars were derived via Monte Carlo simulations with the
	eclipse light curves. }
	\label{ffig6}
\end{figure*}

\begin{figure*}
 \centerline{\psfig{figure=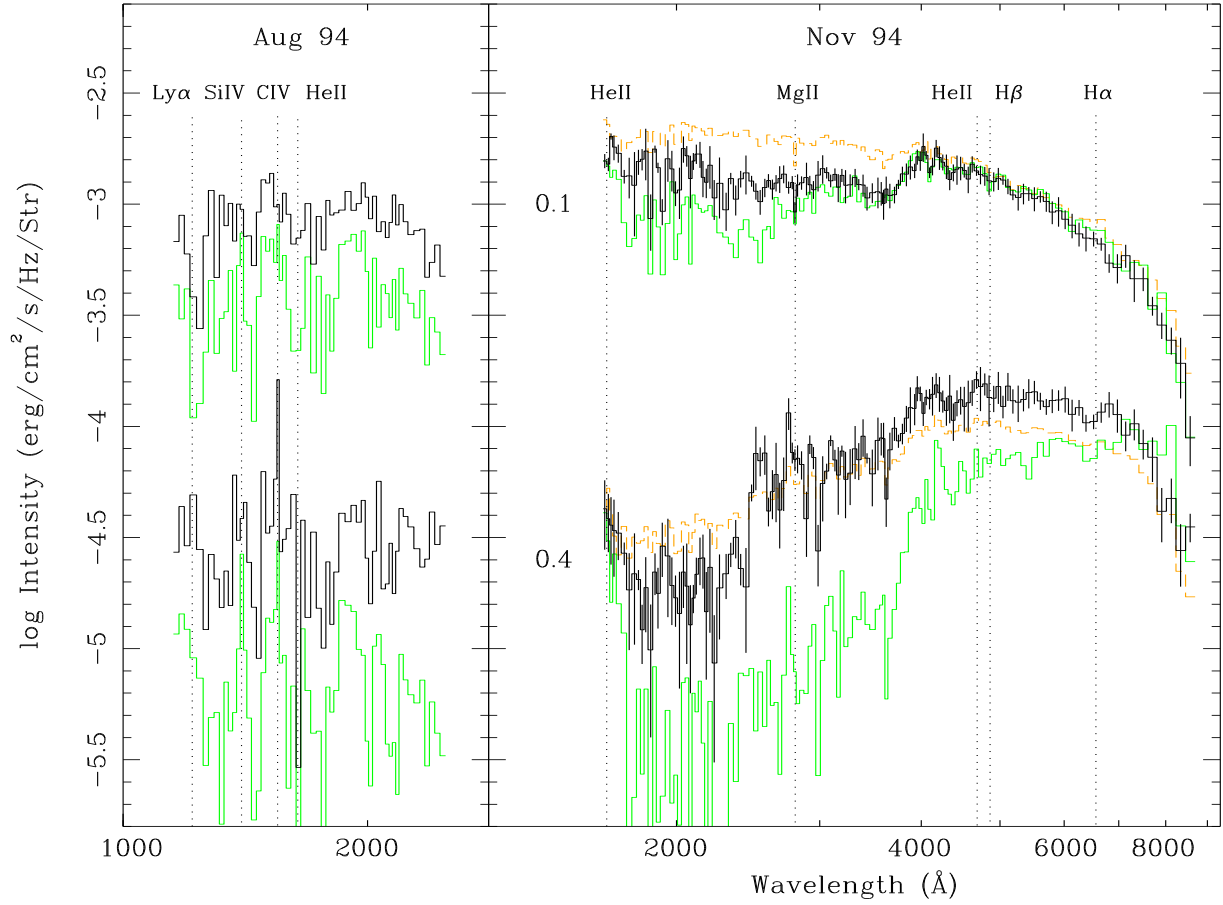,angle=-90,width=18cm,rheight=15cm}}
	\caption{ Comparison of spatially resolved disc spectra for the stream
	(solid black) and disc front (solid gray) regions at two different annuli
	(labels in units of $R_{L1}$). For the Nov 94 data, the spectra of
	the back side of the disc are also shown as dashed gray lines. Error
	bars were computed from Monte Carlo simulations with the light curves.
	To provide a cleaner display, error bars were omitted except for the
	Nov 94 spectra of the stream region. }
	\label{ffig9}
\end{figure*}

The spatially resolved spectra are helpful to constrain the location
of the absorbing gas in UX~UMa. The absorption features become
more pronounced for increasing radius on the front side of the disc
(Fig.\,\ref{ffig8}), but can also be noted in the back side spectrum
at large radii (Fig\,\ref{ffig6}). They also affect the spectrum of the
gas stream region (Fig\,\ref{ffig9}), although to a lesser extent than
in spectra of the front side region at similar radius. These results
suggest the existence of an absorbing ring of cooler gas ($T\sim 8000$\,K)
whose density and/or vertical scale increase with disc radius.
%

Other interpretations deserve comment.  An alternative explanation of
the observed asymmetry is to consider it the result of enhanced front
side limb-darkening effects on a flared disc (e.g., Wade 1996; Robinson,
Wade \& Wood 1998, in preparation).  If the disc has a non negligible
opening angle, the emission from its front side would experience a
stronger limb darkening effect due to its relatively higher viewing
angle and would appear fainter regardless of wavelength. However, the
facts that (1) no front-back asymmetry is observed in the optical and (2)
the absorption by the iron curtain also affects the emission of the back
side of the disc, argue against this interpretation.  For this latter
reason, possible self-obscuration of the front side by a vertically
extended disc rim (as suggested by Smak, in Baptista 1997) is also not
a satisfactory explanation.

Extensive modeling of the spatially resolved spectra
is required in order to properly test these ideas and to provide further
insight into the physical conditions of the UX~UMa accretion disc and
the possible absorbing ring. This is outside of the scope of the present
paper and will be the subject of future work.

\begin{figure*}
 \centerline{\psfig{figure=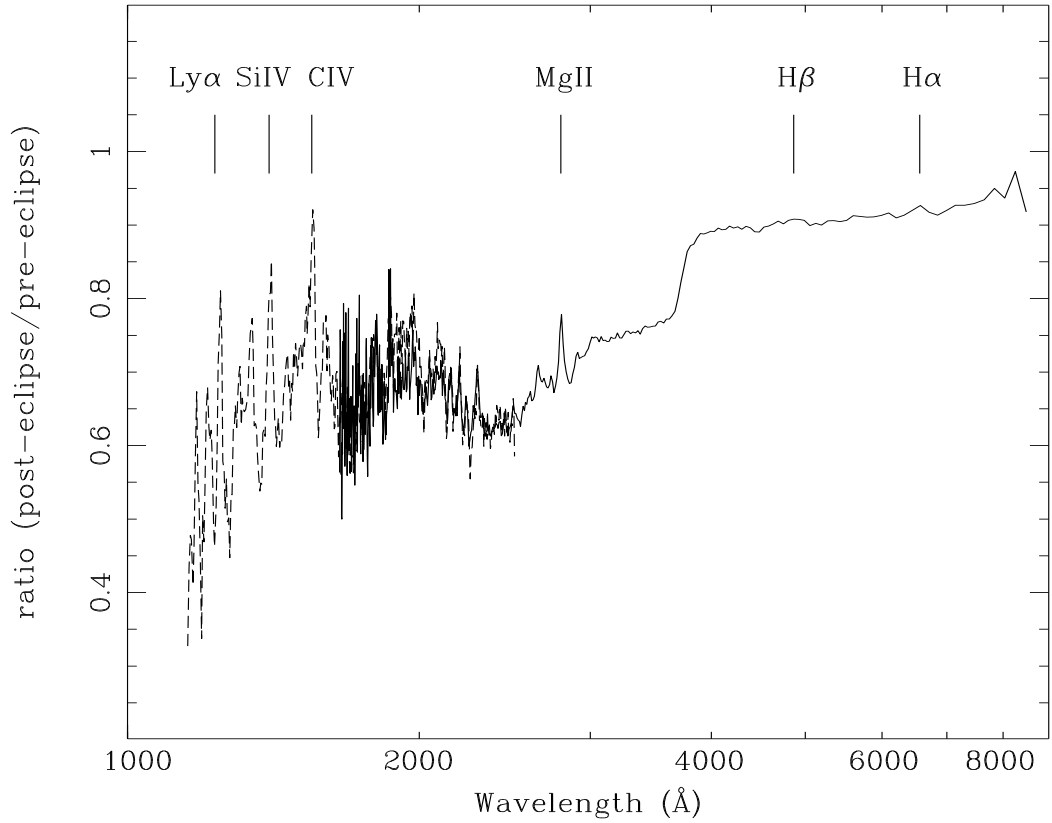,angle=-90,width=18cm,rheight=15cm}}
	\caption{ The ratio of the post-eclipse ($\phi>0.1$) to pre-eclipse
	($\phi<-0.1$) spectra for the G160L (dashed line) and PRISM (solid
	line) data. }
	\label{fignova}
\end{figure*}

Knigge et~al. (1998a) interpreted the noticeable decrease in flux
level of the UV lightcurves from before to after eclipse as due to
anisotropic emission from the bright spot.  Here we give an alternative
interpretation of this effect in terms of phase-dependent absorption
by cool material which is stronger along the line of sight at egress
phases. In this regard, the ratio of the pre- to post-eclipse spectrum
yields a good diagnostic of how much absorption occurs at each wavelength.
Fig.\,\ref{fignova} supports this interpretation by showing that
the ratio of the average post-eclipse ($\phi>0.1$) to pre-eclipse
($\phi<-0.1$) spectra displays the same spectral features characteristic
of absorption by the iron curtain seen in the spatially resolved
spectra of Figs.\,\ref{ffig6} and \ref{ffig8}, as well as in Horne
et~al. (1994). This interpretation is in line with the above remark
that the absorption by the iron curtain in the disc side closest to the
secondary star is more pronounced in the leading quadrant (seen along
the line of sight to the inner disc at post-eclipse phases) than in the
trailing quadrant (which contains the bright spot and gas stream and is
seen along the line of sight to the inner disc at pre-eclipse phases).
Stream impact on the outer edge of the disc (e.g., Hirose, Osaki \&
Mineshige 1991; Meglicki, Wickramasinghe \& Bicknell 1993) or tidal
effects of the secondary star (e.g., Spruit et~al. 1987; Savonije
et~al. 1994) can lead to considerable thickening of the disc rim and
could possibly account for the observed effect.
Alternatively, the absorption may be produced by the material being blown
away in the disc wind. In this case, it may be possible that collision of
the gas stream with the wind `washes' the absorbing gas out of the line
of sight to the accretion disc at orbital phases just before eclipse.
This would result (1) in a relatively smaller absorption effect in the
trailing quadrant and (2) in phase-dependent absorption, which would be
stronger after eclipse than before.
%

\subsection{The gas stream} \label{stream}

Figure\,\ref{ffig9} compares the spectra of the gas stream region with
those of the front side of the disc for a set of annuli. For the
annulus at 0.1 $R_{L1}$, the spectrum of the back side of the disc
is also shown as a dashed gray line. It is seen that the emission
produced in the region containing the gas stream trajectory is different
from the emission of both the front and the back side of the disc.
The gas stream gives a non-negligible contribution to the C\,{\sc IV}
emission, particularly at low (Doppler) velocities, which confirms a
previous suggestion by Baptista et al. (1995).
%
%

\begin{figure*}
 \centerline{\psfig{figure=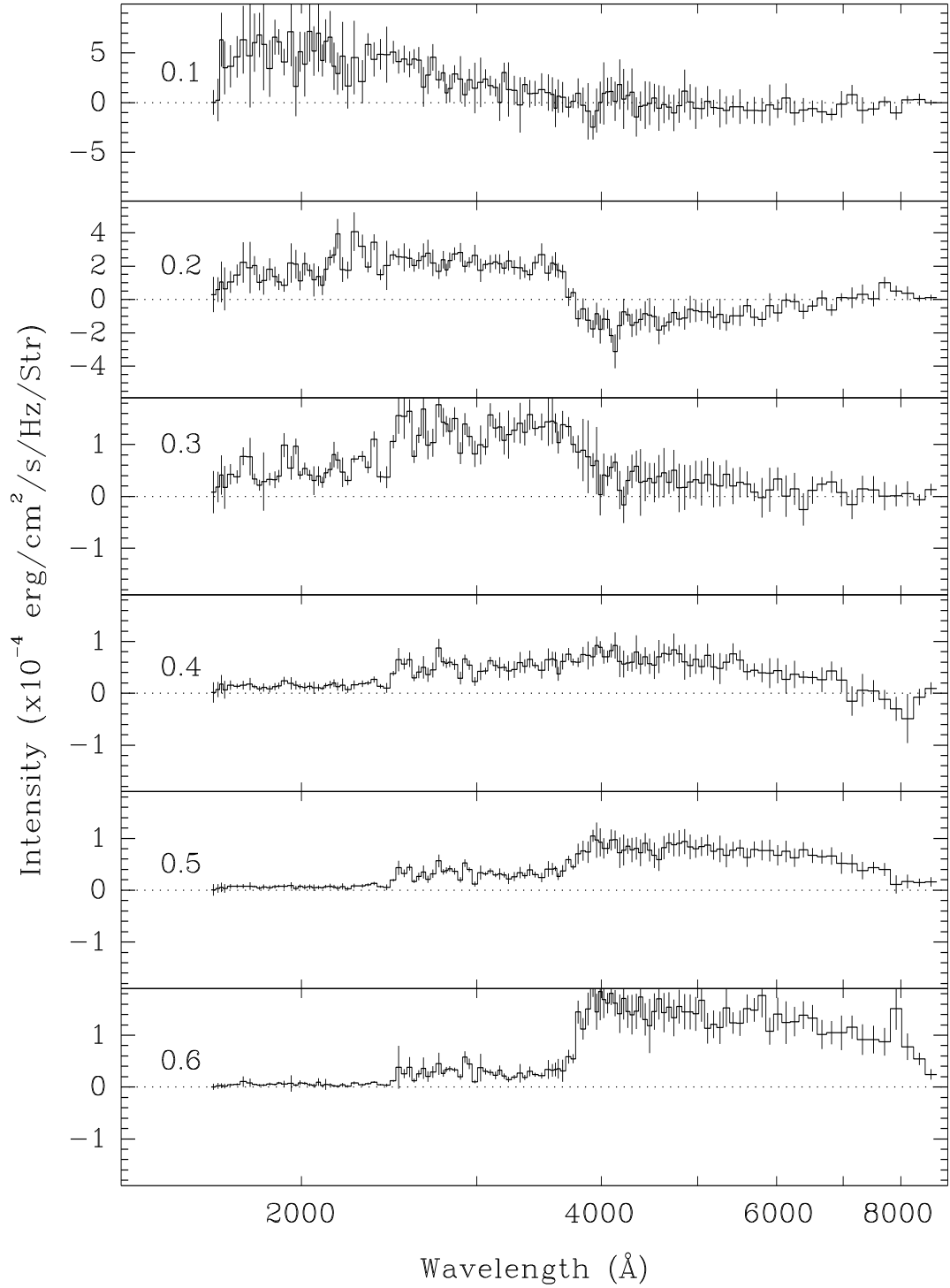,width=17cm,rheight=21cm}}
	\caption{ The difference between the November spectrum of the stream
	and the spectrum of the front side of the disc at same radius for a
	set of annuli (labels in units of $R_{L1}$). }
	\label{ffig10}
\end{figure*}

To gain further insight into the emission properties of this azimuthal
region, we computed the difference between the spectrum of the stream
and that of the front side of the disc as a function of disc radius.
This is shown in Figure\,\ref{ffig10}.
The spectrum of the gas stream is noticeably different from the disc
spectrum at the same radius for a range of radii. The spectrum of the
difference becomes progressively bluer with decreasing disc radius.
This suggest that gas overflows through the impact point at disc rim
and continues along the stream trajectory, producing distinct emission
down to $0.1 \: R_{L1}$.
This effect was not seen in the previous optical mapping experiment
(Rutten et~al. 1994) because the distinct gas stream emission reveals
itself mainly in the ultraviolet region.
%

\subsection{The spectrum of the uneclipsed light} \label{fbg}

The spectrum of the uneclipsed light (Fig\,\ref{ffig6}) shows prominent
emission lines of Ly$\alpha$, N\,{\sc V} $\lambda 1241$, Si\,{\sc IV}
$\lambda 1400$, C\,{\sc IV} $\lambda 1550$, He\,{\sc II} $\lambda 1640$,
and Mg\,{\sc II} $\lambda 2800$ superposed on a ultraviolet continuum
which rises from 0.5 to 5.0 mJy between 1600 and 3600 \AA\ and shows a
roughly constant level of $\simeq 3.5$ mJy in the optical region. The
Balmer jump appears clearly in emission indicating that the uneclipsed
light has an important contribution from optically thin gas.  H$\beta$
and H$\alpha$ are also seen in emission.  These latter lines may be
stronger than they appear since their flux is diluted into the adjacent
continuum due to the large passbands of our data in the optical range.

The fractional contribution of the uneclipsed component to the UV
continuum emission increases towards longer wavelengths, reaching
$\simeq 25$ per cent of the total light at 3600 \AA.  The uneclipsed
light at Ly$\alpha$, N\,{\sc V}, C\,{\sc IV} and Si\,{\sc IV} is even
more significant, reaching about 40\% of the total light.  A substantial
fraction of the light at these lines does not arise from the orbital
plane and is not occulted during eclipse.  The contribution of the
uneclipsed light to the total light at He\,{\sc II} is smaller, at the
level of 15\%.  These results are in accordance with the findings of
Baptista et~al. (1995) and support the scenario in
which these lines originate in a vertically extended disc chromosphere
+ wind, with the He\,{\sc II} emission being more concentrated towards
the orbital plane.  Optically thin emission from this extended region is
probably responsible for the Balmer jump (and lines) in emission observed
in the uneclipsed spectrum.  Support in favor of this argument comes from
the recent detailed modeling of the C\,{\sc IV} wind line of eclipsing
nova-likes by Schlosman, Vitello \& Mauche (1996) and Knigge \& Drew
(1997). Their results suggest the existence of a relatively dense ($n_e
\sim 4 \times 10^{12}$~cm$^{-3}$) and vertically extended chromosphere
between the disk surface and the fast-moving parts of the wind, which
could produce significant amounts of optically thin emission.

An intriguing feature of the spectrum at mid-eclipse (Fig.\,\ref{ffig2})
is that it shows almost no evidence of a Balmer jump.  This can be
explained as the result of optically thin emission from the uneclipsed
disc chromosphere + wind filling in the Balmer jump produced by the
optically thick outer disc which remains visible at these phases.

The comparison of the uneclipsed spectrum of the G160L and PRISM data
reveals that, despite the substantial change in brightness from the August
to the November observations, the uneclipsed continuum light remained
at roughly the same flux level in the overlapping region $\lambda\lambda
1600-2500$ \AA. Thus, the uneclipsed, optically thin continuum emission
seems quite insensitive to changes in mass accretion rate.  On the
other hand, the He\,{\sc II} emission has increased by a factor $\simeq 2$,
in accordance with the relative change in brightness observed at the
inner disc.  This may be an indication that the He\,{\sc II} emission is
connected with the energy release at the inner disc regions.
Baptista et~al. (1995) remarked that the strength of the UV emission
lines seems correlated with the brightness level in UX UMa.  Assuming
that these UV lines are produced in a disc wind, they suggested that
the amount of matter ejected in the wind (and thus the strength of the
emitted lines) is quite sensitive to small fluctuations in the mass
transfer rate (see also Livio 1996).  The observed correlation between
the mass accretion rate and the He\,{\sc II} emission is in line with their
findings and supports their suggestion.

\section{The influence of the assumed geometry on the results}
\label{geo}

\begin{figure*}
 \centerline{\psfig{figure=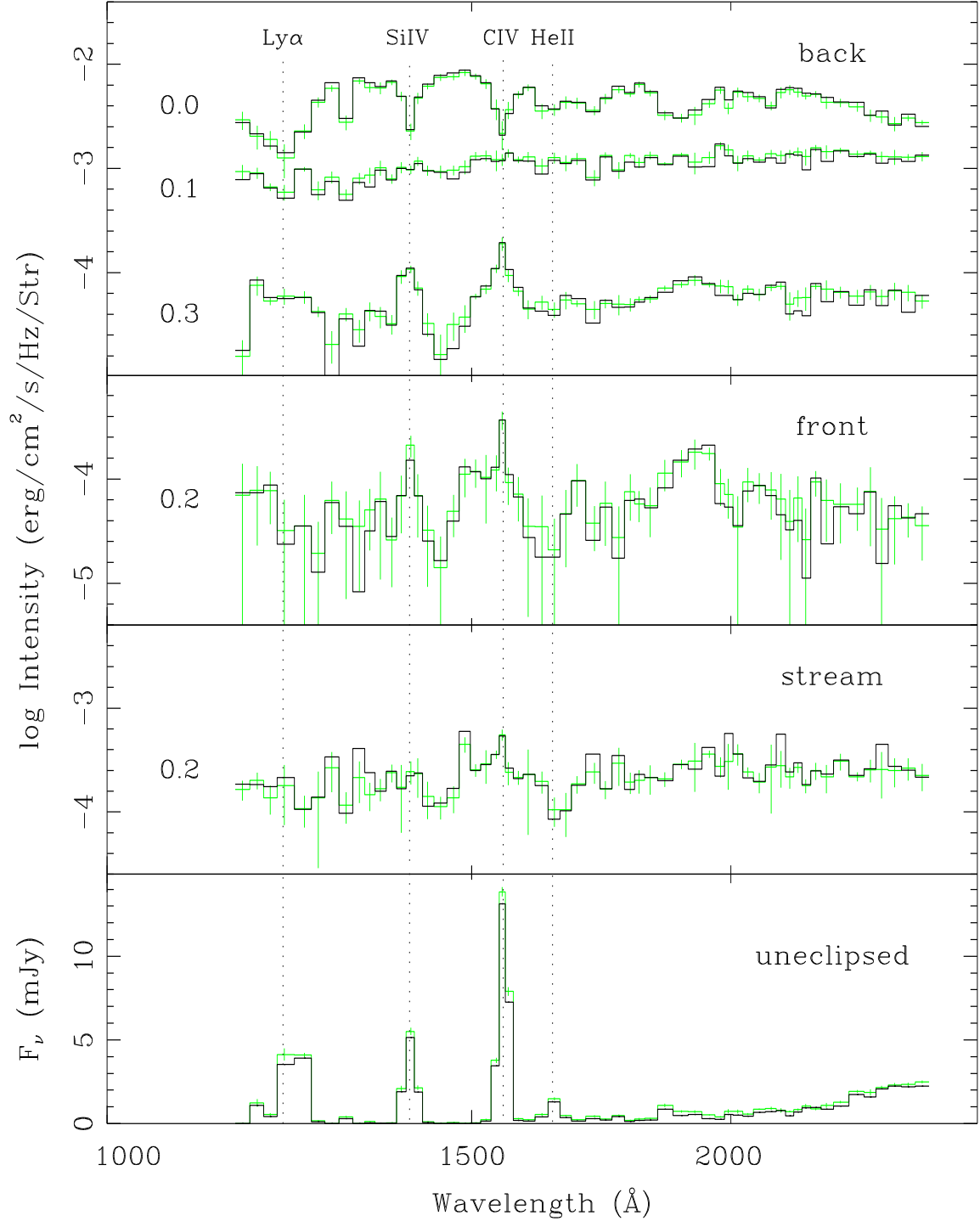,width=17cm,rheight=21cm}}
	\caption{ Comparison of disc spectra at selected radii for
	reconstructions obtained assuming a mass ratio of $q= 1.0$ (gray lines
	with error bars) and $q=0.7$ (solid lines). The lower panel shows the
	spectrum of the uneclipsed component in each case. }
	\label{ffig11}
\end{figure*}

To gauge the sensitivity of the results to the adopted binary
parameters, we produced eclipse maps using the geometry $i=73\degr$
and $q=0.7$, which corresponds to the lower limit at the 3--$\sigma$
level of the nominal solution of Baptista et~al. (1995), and computed
spatially-resolved spectra for the same regions and set of annuli
as before.  Figure\,\ref{ffig11} compares the disc and the uneclipsed
component spectra at selected radii for reconstructions obtained assuming
a mass ratio of $q= 1.0$ with those obtained with $q=0.7$. There are no
systematic differences with wavelength or disc radius and the observed
differences are of the order of the uncertainties at the 1--$\sigma$ level.
We therefore conclude that the results obtained in the previous sections
are reliable and do not depend on possible (systematic) uncertainties
affecting the mass ratio of UX~UMa.
%

\section{Discussion} \label{discuss}

Knigge et~al. (1998a) show that disc models constructed as ensembles of
stellar atmospheres provide poor descriptions of the observed integrated
spectrum of UX~UMa. The disc model spectra are too blue at ultraviolet
wavelengths and overpredict the magnitude of the Balmer jump.  These
problems are not new.  The difficulties in fitting integrated spectra
of nova-likes and dwarf nova in outburst with disc model spectra have a
long history (e.g., Wade 1984, 1988; La Dous 1989; Long et~al. 1991,
1994; Knigge et~al. 1997).  In discussing possible explanations
for these problems, Knigge et~al. (1998a) postulated the presence of a
significant amount of optically thin material in the system in order to
reconcile the disc models with the observed spectrum.

Our spatially resolved study confirms their suggestion by revealing that
the integrated spectrum of UX~UMa has indeed a substantial contribution
from optically thin emission, most probably associated to the uneclipsed
parts of the disc chromosphere + wind.
A calculation by Knigge et~al. (1998a) indicated that the
addition of an optically thin component with $T= 3 \times 10^4$\,K,
$n_H= 5 \times 10^{12}\; {\rm cm}^{-3}$, and vertical extension $H=
9.7 \times 10^9$ cm would be enough to bring the combined disc model
plus optically thin emission into good agreement with the observed PRISM
spectrum.  The predicted fluxes of their optically thin component raises
from $\simeq 2$ mJy at 2000 \AA\ to $\simeq 5.2$ mJy at 3600 \AA, being
at the level of $\simeq 3$ mJy at 4500 \AA\ -- in good accordance with
the fluxes of the uneclipsed component in Fig.\,\ref{ffig6}.  For this
case, their inferred mass accretion rate is $5 \times 10^{17}\; g\,s^{-1}$
or $10^{-8.1}\; M_\odot\: yr^{-1}$, in excellent agreement with our result.

Thus, the reason for the discrepancies between the prediction of the
standard disc model and observations is not an inadequate treatment of
radiative transfer in the disc atmosphere (or standard models of vertical
structure), but rather the presence of additional important sources of
light in the system besides the accretion disc
%
%
(e.g., optically
thin continuum emission from the disc wind and possible absorption
by circumstellar cool gas).  Following the same line of reasoning,
if disc winds are a common characteristic of all non-magnetic, high
state cataclysmic variables, one might expect their disc chromospheres
to contribute a non-negligible amount of optically thin emission to the
total light of the system. Under this hypothesis, the discrepancy between
disc models and the integrated spectrum observed in other non-magnetic,
high state systems may be removed by the inclusion of a proper optically
thin component to their total light.

These results underscore the importance of spatially resolved studies
in disentangling the different components of the integrated spectra of
cataclysmic variables. In this particular case, it helped to clarify the
situation regarding the apparent discrepancy between disc atmospheres
models and the observed spectra.

\section {Conclusions} \label{conclusao}

We used time-resolved spectroscopy obtained with the HST/FOS to
study the structure and the spectra of the accretion disc and
gas stream of the nova-like UX~UMa in the UV and optical ranges.
The main results of this analysis can be summarized as follows:

\begin{itemize}
\item The inner accretion disc of UX~UMa is characterized by a
blue continuum filled with absorption bands and lines, which cross over
to emission with increasing disc radius. The Balmer jump appears clearly in
absorption at intermediate and large disc radii. Together with the deep
absorption lines seen at the inner disc, this suggests that the accretion
disc is everywhere optically thick.

\item The comparison of the front and back side spectra at the same radius
reveals a significant asymmetry in the disc emission at UV wavelengths
which is not seen in the optical. The disc side closest to the secondary
star shows Balmer jump in absorption and pronounced absorption by the
iron curtain.  These results suggest the existence of an absorbing ring of
cold gas whose density and/or vertical scale increase with disc radius.
The substantial reduction in flux level from before to after eclipse
observed in the ultraviolet lightcurves may be the result of phase
dependent absorption by the thick, cool ring.

\item Spatially resolved spectra of the azimuthal region containing
the gas stream suggest that gas overflows through the impact point at
disc rim and continues along the stream trajectory, producing distinct
emission down to $0.1\; R_{L1}$. The distinct stream emission is
only seen at UV wavelengths for $R<0.3 \; R_{L1}$.

\item The spectrum of the uneclipsed light shows prominent emission lines
of Ly$\alpha$, N\,{\sc V} $\lambda 1241$, Si\,{\sc IV} $\lambda 1400$,
C\,{\sc IV} $\lambda 1550$, He\,{\sc II} $\lambda 1640$, and Mg\,{\sc II}
$\lambda 2800$, and a UV continuum rising towards longer wavelengths. The
Balmer jump appears clearly in emission indicating that the uneclipsed
light has an important contribution from optically thin gas. The lines
and the optically thin continuum emission are most probably emitted in
a vertically extended disc chromosphere + wind.

\item The uneclipsed He\,{\sc II} emission varies in accordance with the
increase in brightness from August to November while the UV optically
thin continuum emission seems insensitive to changes in the mass
accretion rate.

\item The radial temperature profiles of the continuum maps are reasonably
well described by a steady-state disc model in the inner and intermediate
disc regions. There is evidence of an increase in the mass accretion
rate from August to November (from \.{M}$= 10^{-8.3\pm 0.1} \;{\rm to}\;
10^{-8.1\pm 0.1}\; M_\odot \; yr^{-1}$), in accordance with the observed
increase in brightness.  Since the UX\,UMa disc seems to be in a high
mass accretion, high-viscosity regime in both epochs, this result suggests
that the {\em mass transfer rate} of UX~UMa varies by a substantial amount
($\simeq 50$ per cent) on time scales of a few months.

\item The reason for the discrepancies between the prediction of the
standard disc model and observations is not an inadequate treatment of
radiative transfer in the disc atmosphere, but rather the presence of
additional important sources of light in the system besides the accretion
disc (e.g., optically thin continuum emission from the disc wind and
possible absorption by circumstellar cool gas).

\end{itemize}

Future work includes fitting disc atmosphere models to the spatially
resolved spectra to derive the distribution of the physical parameters
of the accretion disc -- its temperature, surface density, Mach number,
optical depth, and vertical temperature gradient.  We expect this analysis
to provide useful constraints on current disc atmosphere models and that
this will give new insights into the understanding of accretion discs.

\section*{Acknowledgments}

The spatially resolved spectra of UX~UMa are available in electronic
ascii format upon request to bap@fsc.ufsc.br.  This work was partially
supported by NASA grant GO-5488 from the STScI (which is operated by
AURA under NASA contract NAS\,5-26555) and by CNPq/Brazil research grant
no. 300\,354/96-7.  In addition, RAW acknowledges financial support from
NASA through grant NAG5-3459 and from STScI through grant GO-3683.03,
both to the Pennsylvania State University.  We would also like to thank
Chris Mauche for his contribution to this project.

%

%

\bsp

\end{document}